\documentclass[a4paper]{jpconf}
\usepackage{graphicx}
\usepackage{amsmath}
\usepackage{xcolor}
\newcommand{\beqa}{\begin{eqnarray}}
\newcommand{\eeqa}{\end{eqnarray}}
\newcommand{\ee}{\mathrm{e}}
\newcommand{\ii}{\mathrm{i}}
\newcommand{\eee}{\textrm{e}}
\newcommand{\ggg}{\textrm{g}}
\newcommand{\ppp}{\textrm{p}}
\newcommand{\bra}[1]{\langle #1 |}
\newcommand{\ket}[1]{|#1\rangle}
\newcommand{\braket}[2]{\langle #1 | #2 \rangle}

\begin{document}
\title[Scattering of two photons on a quantum emitter]
{Scattering of two photons on a quantum emitter in a one-dimensional waveguide: Exact dynamics and induced correlations}

\author{Anders Nysteen, Philip Tr{\o}st Kristensen, Dara P.~S. McCutcheon, Per Kaer, and Jesper M{\o}rk}

\address{DTU Fotonik, Department of Photonics Engineering, {\O}�rsteds Plads, 2800 Kgs Lyngby, Denmark}

\ead{anys@fotonik.dtu.dk}

\begin{abstract}
We develop a wavefunction approach to describe 
the scattering of two photons on a quantum emitter embedded in a one-dimensional waveguide.
Our method allows us to calculate the exact dynamics of the complete system at all times, as well as the transmission properties of the emitter. 
We show that the non-linearity of the emitter with respect to incoming photons depends strongly on the emitter excitation and the spectral shape of the incoming pulses, 
resulting in transmission of the photons which depends crucially on their separation and width. In addition, for counter-propagating pulses, we 
analyze the induced level of quantum correlations in the scattered state, 
and we show that the emitter behaves as a non-linear beam-splitter when the spectral width of 
the photon pulses is similar to the emitter decay rate.
\end{abstract}

\section{Introduction}
Single photons play an important role in many of the rapidly emerging quantum technologies~\cite{Obrien09,Claudon10}, including 
quantum communication~\cite{Gisin07}, quantum metrology~\cite{Giovannetti04}, and optical quantum information processing~\cite{Knill01}.
The most ambitious of these technologies require the manipulation of 
data encoded in the state of the photons, necessitating both single and two-photon gates~\cite{Knill01,Kok07,Kok_book}. Whilst single-photon gates 
can be readily implemented with passive linear optical components, 
photons do not inherently interact, and two-photon gates therefore typically require non-linear 
components~\cite{Kok_book}. Owing to the usually weak nature of these non-linearities, their utilisation at the few-photon level represents a significant challenge. 

Significant progress has been made, however, by utilising the relatively strong light--matter interaction between photons and 
semiconductor quantum dots~\cite{Banin99,Bayer00,Santori02}. The idea is to use these nano-structures as `third-parties', 
in order to achieve an effective interaction between two otherwise non-interacting photons. Additionally, quantum dots can be 
placed in various structures to allow for guidance of the incoming and outgoing photons. These setups include 
quantum dots in photonic nanowires~\cite{Claudon10}, 
close to plasmonic waveguides~\cite{Chang07,Chen10}, and 
inside line defects of photonic crystal waveguide slabs (PCWS)~\cite{Lund-Hansen08}. 
The last of these systems opens up the intriguing possibility of all-optical on-chip integrated circuits~\cite{Obrien09,Yao09}, with  
the demonstration of extremely high coupling efficiencies between quantum dots and waveguide modes having recently been achieved~\cite{Arcari14}, 
as too has the precise positioning of quantum dots on substrates thanks to improvements in fabrication techniques~\cite{Surrente09}.

The dynamics of a quantum two-level emitter (TLE) interacting with single-photon wavepackets of infinitely narrow bandwidth 
in a photonic waveguide is well understood. In this scenario, the emitter does not 
become appreciably populated, and the resulting dynamical (Markovian) equations can be solved. Notably, scattering on a resonant TLE 
results in total reflection due to destructive interference between the scattered field and the incoming field on the transmission side of the 
TLE~\cite{Auffeves07,Shen09}. Intrinsic losses, such as phonon coupling and other non-radiative 
processes are known to deteriorate this complete destructive interference effect \cite{Hughes12}, 
as too does decay of the emitter into modes other than the guided mode. 
Deviations are also expected for scattering of non-monochromatic photon pulses when the finite 
width of the incoming pulses is taken into account, resulting in 
 non-zero transmission \cite{Chen11}. 

The scattering of multiple photons with finite bandwidth is much more complicated, as the non-linear emitter 
can induce correlations between the photons caused by elastic multi-photon scattering processes~\cite{Shen07_2,Fan10}. 
Existing methods for analyzing the multiple-photon scattering problem --- such as the input-output formalism~\cite{Fan10}, 
the real-space Bethe ansatz~\cite{Shen07}, or the Lehmann-Symanzik-Zimmermann formalism~\cite{Shi09} --- focus on the long-time limit 
of the scattered state~\cite{Zheng10} and necessitate the computation of complicated scattering elements. 
Some specific considerations have been demonstrated using a wavefunction description of the system~\cite{Kojima03}, 
e.g. the demonstration of stimulated emission of an emitter inside a waveguide~\cite{Valente12}, and 
scattering of a two-photon wavepacket in a photonic tight-binding waveguide~\cite{Longo09,Moeferdt13}. 
Applications which utilize a TLE nonlinearity have been proposed, such as photon sorters and Bell state analyzers~\cite{Witthaut12}. 
In all these cases the non-linearity of the emitter leads to rich scattering dynamics and scattering-induced correlations. 
It is the interplay between these highly non-trivial scattering properties and the excitation dynamics of the emitter which we 
seek to clarify in this work.

To do so we study two-photon scattering on a quantum emitter in a one-dimensional waveguide using a 
wavefunction approach, in which the entire system state is explicitly calculated at all times during the scattering 
process, and which therefore provides a detailed picture of the scattering dynamics. This approach 
relies on a direct solution of the Schr\"odinger equation by 
expanding the complete state in a basis formed by the TLE and the optical waveguide modes. 
This allows us to explore varying widths and separations of the incoming photons, and provides a convenient and 
detailed visualization of the temporal dynamics of the scattering process. As a special case, we show that the 
approach agrees with the above-mentioned methods in the post-scattering limit. For co-propagating pulses, we 
find that the transmission properties of the emitter depend crucially on the pulse width and separation, with closer 
spaced pulses giving rise to a larger proportion of scattered light. For counter-propagating coincident pulses we 
find that the emitter behaves as a non-linear beam-splitter, and we investigate the quantum correlations 
induced in the scattered photonic state. 

\begin{figure}
\begin{center}
\includegraphics[width=0.85\textwidth]{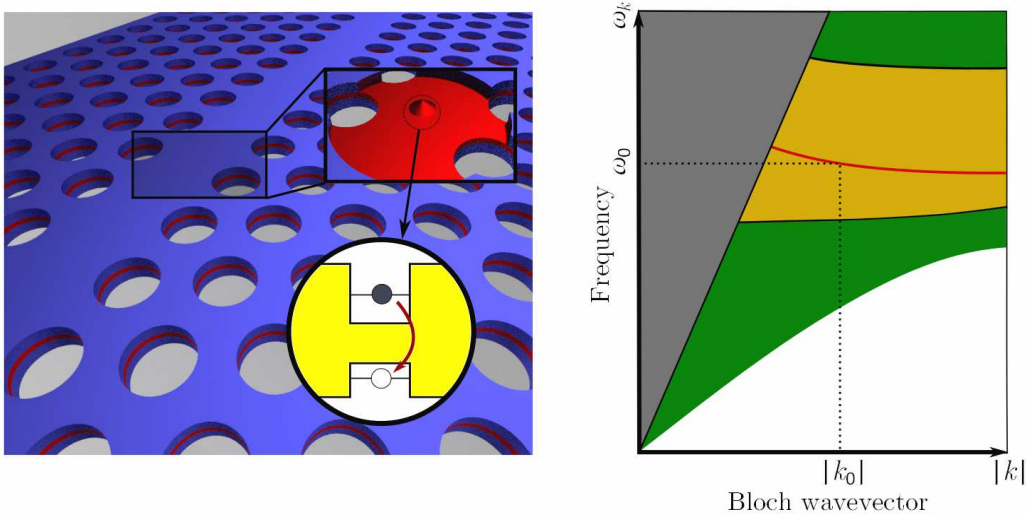}
\caption{\label{fig_sketch} (Left) Illustration of a TLE embedded in a one-dimensional waveguide, 
exemplified by a line defect in a photonic crystal slab containing a quantum dot. 
(Right) Schematic illustration of the corresponding band diagram showing the slab modes (green area) with a bandgap 
(yellow area) containing a line defect mode (red line). The resonance frequency of the emitter, 
$\omega_0$, lies inside the bandgap, and we consider only propagating modes below the light cone (shaded grey).}
\end{center}
\end{figure}

This paper is structured as follows: In section~\ref{model} we introduce the model and formalism. 
In section~\ref{results} we analyse the scattering dynamics 
for two co-propagating photon pulses; we examine how the properties of the scattered state depend 
on the emitter excitation and consider the scattering-induced correlations between the photons. 
In section~\ref{counter} we study scattering of counter-propagating pulses, elucidating the analogy 
of the quantum emitter and a non-linear beam-splitter. Finally, in section~\ref{conclusion} we summarise our results.

\section{The model}
\label{model}
The model we study consists of a TLE coupled to an infinite one-dimensional waveguide with modes propagating in both directions. 
This model could be realized, for example, by a line defect in a photonic crystal containing a quantum dot, 
as depicted in Fig.~{\ref{fig_sketch}}. The complete Schr\"{o}dinger picture Hamiltonian reads $H=H_0+H_I$, 
where $H_0=\hbar\omega_0 c^{\dagger} c + \sum_\lambda\hbar\omega_\lambda a_\lambda^\dagger a_\lambda$ 
and $H_I=\hbar\sum_\lambda [ g_\lambda a_\lambda c^\dagger+ g_\lambda^* a_\lambda^\dagger c ]$, 
in which $\lambda$ is a generalised quantum number describing polarisation and propagation degrees of freedom, 
and each mode is described by creation and annihilation 
operators $a^{\dagger}_{\lambda}$ and $a_{\lambda}$, respectively. The TLE is 
described by creation and annihilation operators $c^{\dagger}$ and $c$ and 
has excited state energy $\hbar\omega_0$. The coupling between the TLE and mode $\lambda$ is $g_\lambda$.

Moving into a rotating frame described by the transformation 
$T(t)=\exp[-\ii \omega_0 (c^{\dagger} c +\sum_{\lambda} a_{\lambda}^{\dagger}a_{\lambda})]$, we find the 
transformed Hamiltonian $\tilde{H}=T^{\dagger}(t) H T(t)+\ii \hbar \frac{\partial T^{\dagger}}{\partial t} T(t)=
\tilde{H}_0+\tilde{H_I}$ where 
\beqa
\tilde{H}_0=\sum_\lambda\hbar\Delta\omega_\lambda a_\lambda^\dagger a_\lambda, \qquad\text{and}
\qquad
\tilde{H}_I=H_I=\hbar\sum_\lambda\left[ g_\lambda a_\lambda c^\dagger+ g_\lambda^* a_\lambda^\dagger c\right],
\label{two_fot_Ham}
\eeqa
where $\Delta\omega_{\lambda}=\omega_{\lambda}-\omega_0$ is the detuning of mode $\lambda$ from 
the TLE emitter transition energy. From this point onwards we work exclusively in this rotating frame.

\subsection{Dynamics}

For photonic applications, the TLE would ideally couple exclusively to guided modes in the waveguide, leading to a lossless system in which the number of excitations is strictly conserved. We note that recent experimental work has shown coupling of a quantum dot  
to modes in a one-dimensional waveguide with an efficiency of up to $98$\%~\cite{Arcari14}. In our analysis we therefore assume coupling only to the waveguide modes, which allows us to expand a general state of the system in a basis spanned by the states $\ket{\ggg,\lambda_1\lambda_2}$ 
and $\ket{\eee,\lambda}$, where the first index refers to the TLE in the ground (g) or excited (e) state, and the second index labels the population of the waveguide mode(s). We note that since the photons are fundamentally indistinguishable, the states $\ket{\ggg,\lambda_1\lambda_2}$ and 
$\ket{\ggg,\lambda_2\lambda_1}$ are equivalent. 

We write the total state at time $t$ as
\beqa
\ket{\psi(t)}=\frac{1}{\sqrt{2}}\sum_{\lambda_1\lambda_2}C^\ggg_{\lambda_1\lambda_2}(t)a^\dagger_{\lambda_1}a^\dagger_{\lambda_2}\ket{\ggg,0}
+\sum_\lambda C^\eee_{\lambda}(t)a^\dagger_{\lambda}\ket{\eee,0}, \label{twofot_1}
\eeqa
where the expansion coefficients $C^\ggg_{\lambda_1\lambda_2}(t)$ and $C^\eee_{\lambda}(t)$ are in the rotating frame, and 
$|0\rangle$ indicates the vacuum state of the waveguide. Since $[a^\dagger_{\lambda_1},a^\dagger_{\lambda_2}]=0$ 
(and indeed $\ket{\ggg,\lambda_1\lambda_2}=\ket{\ggg,\lambda_2\lambda_1}$), 
the coefficients of the two-photon terms must be symmetric, $C^\ggg_{\lambda_1\lambda_2}(t)=C^\ggg_{\lambda_2\lambda_1}(t)$. 
Normalization of the state requires 
$\braket{\psi(t)}{\psi(t)}=\sum_{\lambda_1,\lambda_2}|C^\ggg_{\lambda_1\lambda_2}(t)|^2+\sum_\lambda|C_\lambda^\eee(t)|^2=1$, 
and we can interpret $\sum_\lambda|C_\lambda^\eee(t)|^2$ as the probability that the TLE is measured in its excited state, while 
the probability of measuring two photons in modes $\lambda_1$ and $\lambda_2$ 
for $\lambda_1\neq \lambda_2$ is $2|C^\ggg_{\lambda_1\lambda_2}(t)|^2$, and $|C^\ggg_{\lambda_1\lambda_1}(t)|^2$ for $\lambda_1=\lambda_2$. 
Inserting Eq.~({\ref{twofot_1}}) into the time-dependent Schr\"odinger equation, and 
using the Hamiltonian in Eq.~(\ref{two_fot_Ham}) leads to a system of coupled differential equations for the expansion coefficients: 
\begin{subequations}
\begin{eqnarray}
\partial_t C^\eee_\lambda(t) &=& -\ii\Delta\omega_\lambda C_\lambda^\eee(t)-\ii\sqrt{2}\sum_{\lambda'}g_{\lambda'}C^g_{\lambda\lambda'}(t), \\
\partial_t C^\ggg_{\lambda_1\lambda_2}(t) &=& -\ii(\Delta\omega_{\lambda_1}+\Delta\omega_{\lambda_2})C^\ggg_{\lambda_1\lambda_2}(t)-\frac{\ii}{\sqrt{2}}\Big(g_{\lambda_1}^*C^\eee_{\lambda_2}(t)+g_{\lambda_2}^*C^\eee_{\lambda_1}(t)\Big).
\label{diff_lambda}
\end{eqnarray}
\end{subequations}

For a one-dimensional waveguide, such as the photonic crystal line defect in Fig. \ref{fig_sketch}, 
it is reasonable to choose a frequency span where the emitter couples to only two 
waveguide modes propagating in opposite directions. In this case, the mode index $\lambda$ labels modes described by a wavenumber $k$, 
having only a single polarisation, and where a positive or negative value of $k$ implies a waveguide mode 
propagating to the right or left, respectively. 
With these assumptions, the sum over all modes in the waveguide reduces to 
$\sum_ \lambda=\lim_{L\rightarrow \infty}(L/2\pi)\int_{-\infty}^\infty\textrm{d}k\,$, with $L$ being the length of the 1D waveguide, 
and $2\pi/L$ the spacing between the modes in reciprocal space. 
By defining continuous mode versions of the discrete functions and variables in 
Eqs.~(\ref{diff_lambda}), $C^\ggg(k,k',t)=\lim_{L\rightarrow \infty}(L/2\pi)C^\ggg_{\lambda\lambda'}(t)$, 
$C^\eee(k,t)=\lim_{L\rightarrow \infty}\sqrt{(L/2\pi)}C^\eee_\lambda(t)$, 
$g(k)=\lim_{L\rightarrow\infty}\sqrt{(L/2\pi)}g_\lambda$, $\Delta\omega(k)=\Delta\omega_\lambda$, we have 
\begin{subequations}
\begin{eqnarray}
\partial_t C^\eee(k,t)&=&-\ii\Delta\omega(k)C^\eee(k,t)-\ii\sqrt{2}\int_{-\infty}^\infty\textrm{d}k'\,g(k')C^\ggg(k,k',t), \label{diff_lign_int1}\\
 \partial_t C^\ggg(k_1,k_2,t)&=&-\ii[\Delta\omega(k_1)+\Delta\omega(k_2)]C^\ggg(k_1,k_2,t)\nonumber\\
&&\;\;\;\;\;\;\;\;\;\;\;\;\;\;\;\;\;\;\;\;-\frac{\ii}{\sqrt{2}}\big[g^*(k_1)C^\eee(k_2,t)+g^*(k_2)C^\eee(k_1,t)\big].
\label{diff_lign_int2}
\end{eqnarray}
\end{subequations}
By discretizing the $k$-continuum of modes, Eqs.~(\ref{diff_lign_int1}) and ({\ref{diff_lign_int2}}) constitute a system of coupled linear 
differential equations; for certain input pulse shapes they can be solved analytically~\cite{Valente12}, 
but in general we solve them numerically. We note that in contrast to the linear nature of Eqs.~(\ref{diff_lign_int1}) and ({\ref{diff_lign_int2}}), the Heisenberg equations of motion for the system operators used in the scattering matrix approach result in a set of coupled nonlinear differential equations, 
whose solution must instead be obtained using, e.g. the input--output formalism~\cite{Fan10}.

Within the Wigner-Weisskopf theory, the spontaneous emission rate of the emitter is 
given by $\Gamma=\sum_\lambda 2\pi |g_\lambda|^2\delta(\omega_\lambda-\omega_0)$, 
and is typically of the order $\sim 10^9-10^{10}~\mathrm{s}^{-1}$ for quantum dots~\cite{Stobbe09}; 
this is much less than the optical carrier frequencies of the pulses, which are typically of the order $\omega_0\sim 10^{15}~\mathrm{s}^{-1}$.  
Furthermore, by assuming a smooth dispersion curve for the waveguide modes, e.g. as shown in Fig. \ref{fig_sketch}, 
the waveguide dispersion may be linearized around $\omega_0$, giving 
$\Delta\omega(k)\approx v_{\ggg}(|k|-k_0)$ with the group velocity 
$v_{\ggg}=(\partial\omega/\partial k)|_{k=k_0}$. 

\subsection{Two-photon input state}
Eqs.~(\ref{diff_lign_int1}) and (\ref{diff_lign_int2}) can in principle be solved for any initial state of the total system containing two excitations. 
The case of a single exponentially shaped pulse scattering on an already excited TLE has been considered using a similar 
approach in Refs.~\cite{Valente12,Rephaeli12}. We build on these results by 
considering two optical pulses in the initial state, and investigate their scattering on the 
TLE for various pulse widths and separations. The two-photon input states can be 
experimentally produced using, for example,
parametric down-conversion, as has been demonstrated~\cite{Cinelli04,Ostermeyer09,Harder13}. 
In general, such a process creates two correlated photons, but 
the properties of the down-conversion crystal can be modified in such a way that uncorrelated photons are produced~\cite{Grice01}. 

We write the general form of a two-photon state as 
\beqa
\ket{\beta}=\frac{1}{\sqrt{2}}\int_{-\infty}^\infty \textrm{d}k\,\int_{-\infty}^\infty \textrm{d}k'\,\beta(k,k')a^\dagger(k)a^\dagger(k')\ket{0},
\label{twophot_state}
\eeqa
with $\beta(k,k')$ the two-photon wavepacket given in two-dimensional $k$-space. 
The bosonic nature of the photons implies symmetry of the two-photon wavepacket, $\beta(k,k')=\beta(k',k)$, and the normalisation condition is then  
$\braket{\beta}{\beta}=\int_{-\infty}^\infty \textrm{d}k\,\int_{-\infty}^\infty \textrm{d}k'\,|{\beta}(k,k')|^2=1$. 
If we assume an initial condition corresponding to two photons described 
by Eq.~({\ref{twophot_state}}), 
by comparison with Eq. (\ref{twofot_1}) we find the corresponding initial conditions 
for the wavefunction coefficients $C^\eee(k,0)=0$ and $C^\ggg(k,k',0)={\beta}(k,k')$. 

We write a general symmetric two-photon Gaussian state as $\beta(k,k')=K\left[\beta_0(k,k')+\beta_0(k',k)\right]$ with 
\beqa
\beta_0(k,k')=f\big(k-k_{\ppp,1}+k'-k_{\ppp,2}\big)\xi_1(k)\xi_2(k'),\label{eq:betazz}
\eeqa 
where $\xi_i(k)=\sigma_i^{-1/2}\pi^{-1/4}\exp\left[-\ii z_{0,i} (k-k_{\ppp,i})-(k-k_{\ppp,i})^2/(2\sigma_i^2)\right]$ is a Gaussian single-photon 
wavepacket with $\sigma_i$ describing the spectral width, $z_{0,i}$ the initial position of the pulse center, and where 
positive and negative $k_{\ppp,1}$ or $k_{\ppp,2}$ correspond 
to wavepackets propagating to the right and left respectively. $K$ is a normalization parameter, and $f(k,k')$ is a function 
describing phase matching, which for simplicity may be assumed to be a Gaussian, 
$f(k)=\exp\left[-k^2/2\sigma_p^2\right]$~\cite{Wang06}. The correlation between the two photons is 
described by the parameter $\sigma_p$, which for parametric down-converted photons corresponds to the 
bandwidth of the pump laser~\cite{Wang06}. The correlation 
parameter $\sigma_p$ is inversely proportional to the correlation length between the photons, and thus 
$\sigma_p\rightarrow \infty$ corresponds to fully uncorrelated photons, and in such a 
case $\beta_0(k,k')$ factorizes into two single-photon wavepackets. 
We also define the real-space representation of the two-photon wave-packet by the Fourier transform
\beqa
\beta(z,z')=\frac{1}{2\pi}\int_{-\infty}^\infty\textrm{d}k\int_{-\infty}^\infty\textrm{d}k'\,\beta(k,k')\ee^{\ii kz+\ii k' z'}.
\eeqa   
 
\begin{figure}
\begin{flushright}
\includegraphics[width=1\textwidth]{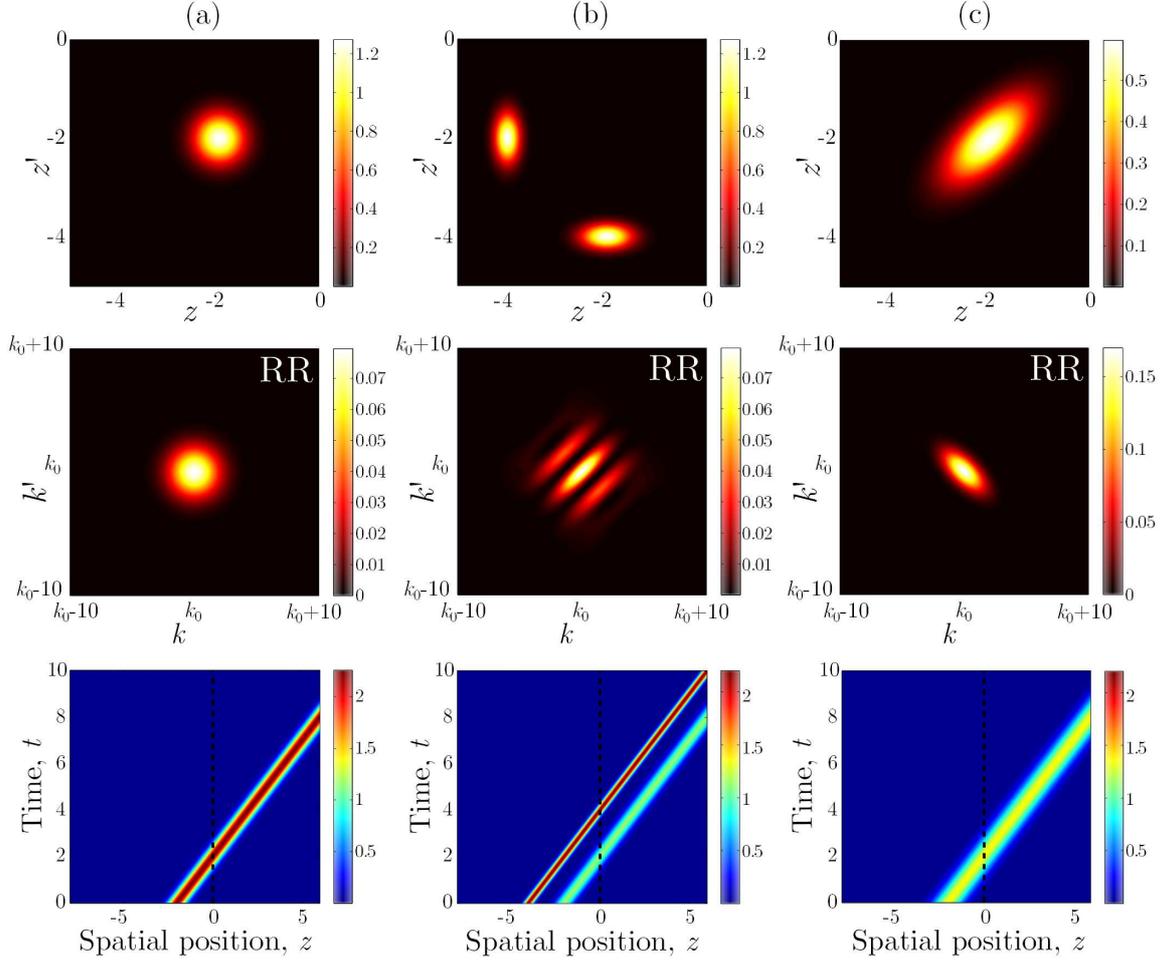}
\caption{\label{rho_plots} 
Absolute value of the two-photon wavepacket in real space, $|\beta(z,z')|$ (upper row), in reciprocal 
space, $|\beta(k,k')|$ (middle row), and the photon density $N(z,t)$ (lower row) for three different two-photon states, and 
with no emitter positioned in the waveguide. The three columns correspond 
to initial photonic states which are co-propagating coincident uncorrelated pulses of equal width ($\sigma_1=\sigma_2=2$, $z_{0,1}=z'_{0,2}=-2$, and 
$\sigma_\textrm{p}\rightarrow \infty$, column (a)), uncorrelated spatially separated pulses of unequal widths 
($\sigma_1=2$, $\sigma_2=4$, $z_{0,1}=-2$, $z'_{0,2}=-4$, and $\sigma_\textrm{p}\rightarrow \infty$, column (b)), 
and coincident highly correlated pulses of equal width ($\sigma_1=\sigma_2=2$, $z_{0,1}=z'_{0,1}=-2$, 
and $\sigma_\textrm{p}=(3/4)\sigma_1$, column (c)).
}
\end{flushright}
\end{figure}
 
In addition to the two-photon wavepacket, described by the functions $\beta(k,k')$ and $\beta(z,z')$, 
it is also useful to define the expectation value of the photon density at a time $t$ and position 
$z$ as $N(z,t)=\bra{\psi(t)}a^\dagger(z)a(z)\ket{\psi(t)}$, where $a(z)=(2\pi)^{-1/2}\int\textrm{d}k\,a(k)\ee^{\ii k z}$ annihilates 
a photon at position $z$. This function has units of $\mathrm{m}^{-1}$, and describes 
the distribution of energy in the waveguide.  In terms of the wavefunction coefficients its explicit form is given by
\beqa
N(z,t)=2\int^\infty_{-\infty}\textrm{d}k\,\bigg|\frac{1}{\sqrt{2\pi}}\int^\infty_{-\infty}\textrm{d}k'\,C^\ggg(k,k',t)\ee^{\ii k'z}\bigg|^2 + \bigg|\frac{1}{\sqrt{2\pi}}\int^\infty_{-\infty}\textrm{d}k\,C^\eee(k,t)\ee^{\ii kz}\bigg|^2, \label{twofot_density} 
\eeqa
and since a lossless system is assumed, the number of excitations is conserved and we find $\int^\infty_{-\infty}\textrm{d}z\,N(z,t)=2$ at all times.

To gain some intuition as to how these three descriptions of the two-photon state appear, we first consider 
three different input states in the waveguide containing no TLE (such that wavepackets propagate along 
the waveguide but no other dynamics are present). The three rows in Fig.~{\ref{rho_plots}} correspond 
to the absolute value of the initial real-space photon wavepacket $|\beta(z,z')|$, the initial $k$-space wavepacket $|\beta(k,k')|$, and 
the photon density as a function of time $N(z,t)$, for input states which correspond to two coincident uncorrelated photons of equal width (column a), 
two spatially separated uncorrelated photons of different width (column b), and two coincident highly-correlated photons (column c). We note that in comparing 
columns (a) and (b), the separated nature of the two pulses in (b) is clearly visible, as too is the inequality of the two pulse widths, 
as is evident from the elliptical shape of the wavepactet amplitudes in the top row. We also see oscillations appearing the $k$-space representation 
for the spatially separated pulses in column (b). These oscillations have a period $|z_{0,1}-z_{0,2}|^{-1}$ and are a signature of 
interference between the two separated pulses. For the correlated pulses in column (c) we see that the wavepacket is 
elongated along the diagonal line $z=z'$ in real-space, and along the $k=-k'$ direction in frequency space. This means 
that position measurements of the two photons will share positive correlations, whereas frequency measurements will be 
anti-correlated. Finally, we note that the photon density plots in the lower row provide us with an overall picture 
of the dynamics for all times, but do not capture all the features present in the photon wavepackets.

\section{Co-propagating pulses}
\label{results}
We now turn to the main focus of this work, and consider the evolution of the two photon state as 
it scatters on a TLE placed inside the waveguide. In order to solve Eqs.~(\ref{diff_lign_int1}) and 
(\ref{diff_lign_int2}), we discretize the continuum of waveguide modes and numerically solve the resulting finite set of differential equations. 
In the following calculations we assume frequency-independent coupling constants, $g(k)=g$, 
which is well justified owing to the assumption that the TLE linewidth is narrow compared to the 
carrier frequency of the wavepackets; in general, the 
approach we use allows for frequency dependent coupling constants, which could be relevant, for example, 
when considering coupling to optical cavities~\cite{Vahala03}. Convergence tests were performed by 
comparison with the well-known scattering properties of a single photon~\cite{Chen11}, and to analytical expressions for the induced TLE excitation probability 
obtained by solving Eqs.~(\ref{diff_lign_int1}) and (\ref{diff_lign_int2}) for a 
two-photon Gaussian input pulse (using the method of Refs.~\cite{Kojima03,Valente12}). Our results in the long-time limit agree with the scattering matrix 
approach of Refs.~\cite{Fan10,Zheng10}. In all plots, parameters with units of time or length are 
normalized to $\Gamma^{-1}$ and $v_{\ggg}/\Gamma$ respectively. A pulse with a spectral width of $\sigma=1$ 
thus corresponds to a spatial width of $v_{\ggg}/\Gamma$ and a temporal width of $\Gamma^{-1}$. Finally, 
for plotting in $z$-space, we used a frequency of $\omega_0=10^{15}~\mathrm{s}^{-1}$. 

\subsection{Scattering dynamics}
\begin{figure}
\begin{flushright}
\includegraphics[width=1\textwidth]{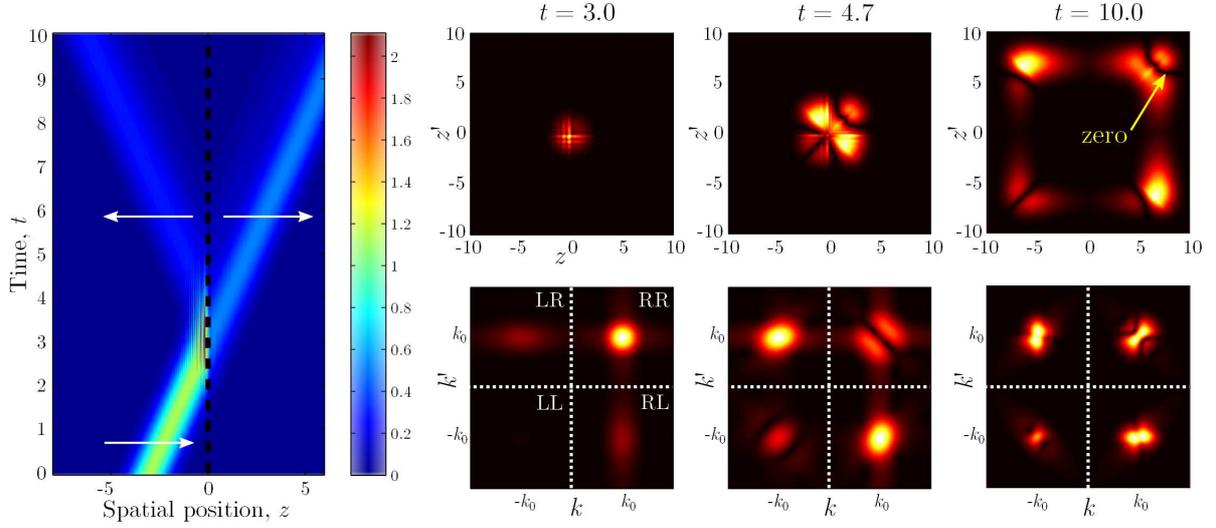}
\caption{\label{sameSide_dynamics_coin} 
Left: Photon density, $N(z,t)$ for an initially uncorrelated ($\sigma_\textrm{p}\rightarrow \infty$) coincident 
two-photon state scattering on an emitter placed at $z=0$, using widths $\sigma_1 =\sigma_2=1$ and initial centre positions 
$z_{0,1}=z_{0,2}=-3$. The position of the emitter at $z=0$ is indicated by the black dashed line.
Right: Absolute value of the two-photon wavepacket shown at three representative  
times during the scattering event, both in $z$-space (upper row) and $k$-space (lower row). 
In the $k$-space plots, we show only the regions centred around $k,k'=\pm k_0$, which 
we label $LL$ (origin $(-k_0,-k_0)$), $RR$ (origin $(k_0,k_0)$), $LR$ (origin $(-k_0,k_0)$), and $RL$ (origin $(k_0,-k_0)$).}
\end{flushright}
\end{figure}

As an illustrative example of two-photon scattering, we first consider the scattering of two identical, 
coincident and uncorrelated single-photon pulses with carrier frequencies resonant with the TLE. Except for the inclusion of a TLE here, 
the input is identical to that of column (a) in Fig.~\ref{rho_plots}; 
both photons are initially located left of the TLE, $z_{0,1}=z_{0,2}<0$, and propagate to the right, $k_{p,1}=k_{p,2}>0$. 
On the left of Fig.~\ref{sameSide_dynamics_coin} we show the photon density $N(z,t)$, which represents the expectation 
value of position measurements of the two photons over many scattering events. We see 
that part of the energy is transmitted, and part is reflected. On the incoming side of the emitter ($z<0$), 
a standing wave pattern is clearly visible, which is a result of interference between the incoming and reflected part of the pulse. 

The upper row on the right shows the evolution of the spatial wavepacket at three representative 
times, corresponding to the onset of the scattering $t=3.0$, during the scattering $t=4.7$, 
and in the post-scattering long-time limit $t=10.0$. We notice that after the scattering event, 
both photons clearly propagate away from the TLE as expected. Additionally, 
the scattered state contains all possible spatial configurations of the photons: both being in the region right of the TLE, ``$RR$", 
one on each side, ``$LR$", and both photons to the left of the TLE, ``$LL$". 
An equivalent conclusion may also be drawn from the wavepacket in $k$-space as shown in the 
lower row of~Fig. \ref{sameSide_dynamics_coin}, where the scattered field has components propagating 
in the ``$RR$", ``$LR$", or ``$LL$" directions. Due to the bosonic nature of the photons, 
the configurations ``$LR$" and ``$RL$" cannot be distinguished. 
At early times, e.g. at $t=3.0$ in Fig.~\ref{sameSide_dynamics_coin}, the scattering is dominated by single photon processes, which can be seen by the fact that the two-photon wavepacket is elongated along the $k$ and $k'$ axes. This means that only a single photon has been broadened by its interaction with the TLE emitter, whilst the other remains unchanged. At larger times, features of two-photon scattering processes appear, which can be seen by the more complex shapes of the two-photon wavepackets. We discuss these features in more detail below.

\begin{figure}
\begin{flushright}
\includegraphics[width=1\textwidth]{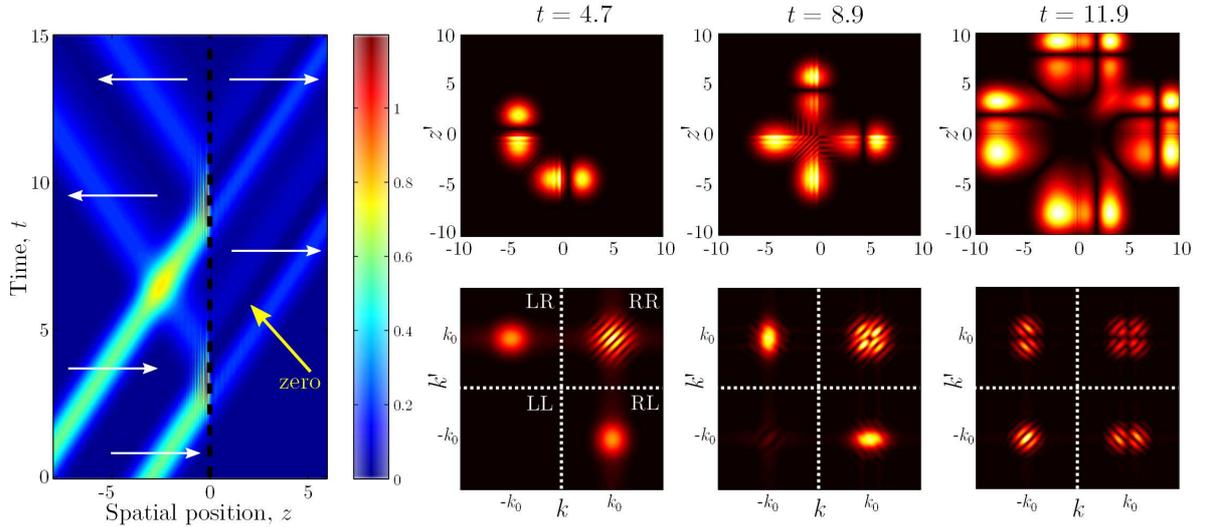}
\caption{\label{sameSide_dynamics_sep}
Left: Photon density, $N(z,t)$ for an uncorrelated ($\sigma_\textrm{p}\rightarrow \infty$) 
two-photon state scattering on the emitter placed at $z=0$, using widths $\sigma_1 =\sigma_2=1$ and initial centre positions 
$z_{0,1}=-3$ and $z_{0,2}=-9$. The position of the emitter at $z=0$ is indicated by the black dashed line.
Right: Absolute value of the two-photon wavefunction shown at three different times during the scattering event, 
both in $z$-space (upper row) and $k$-space (lower row). In the $k$-plots, only the 
regions centred around $k,k'=\pm k_0$ are shown.
}
\end{flushright}
\end{figure}

It is interesting to compare the scattering dynamics in Fig.~\ref{sameSide_dynamics_coin} with the case 
of two pulses which are sufficiently separated in space such that the TLE excitation induced by the first pulse has essentially 
decayed before the arrival of the second pulse. 
This is shown in Fig.~\ref{sameSide_dynamics_sep}, and in this case the scattering behaviour resembles two 
`copies' of the single-photon scattering case~\cite{Chen11}. Even though the carrier frequency of the pulse is 
resonant with the TLE, a non-zero transmission is obtained in this single-photon scattering limit because of the finite temporal widths of the input pulses. 
These features are in contrast to the case in which a spectrally narrow continuous wave pulse is incident on the emitter, 
which gives zero transmission on resonance because of destructive interference between the scattered and input 
fields~\cite{Auffeves07,Shen09}. In this single-photon scattering limit, the TLE fully reflects frequency components of 
the incoming pulse which are close to the TLE resonance, as no two-photon processes are apparent. Hence, the 
spectrum of the transmitted pulse does not contain components at these frequencies, see e.g. the spectrum in 
Fig.~\ref{sameSide_dynamics_sep} at $t=11.9$. This is in contrast to 
the coincident case in Fig.~\ref{sameSide_dynamics_coin}, where two-photon processes allow for 
transmission of pulse components close to the TLE resonance.

During the initial phase of the scattering, the $k$-space wavefunctions in both Fig.~\ref{sameSide_dynamics_coin} and 
Fig.~\ref{sameSide_dynamics_sep} broaden and demonstrate interaction with states which are detuned from the TLE by several TLE linewidths. 
This may be seen at times $t=3.0$ and $t=4.7$ in Fig.~\ref{sameSide_dynamics_coin}, but these frequencies do 
not appear in the final scattered state at $t=10.0$. This phenomenon may be understood as arising from the 
energy-time uncertainty relation, as processes taking place 
at short times allow for larger uncertainties in energy. Lastly, for the case of spatially separated pulses in 
Fig.~\ref{sameSide_dynamics_sep}, a dip is present in the transmitted waveguide excitation. 
This feature is a consequence of destructive interference between the initial photon wavepacket 
and the emitted photon, and manifests in the form of a dip in the spectrum of the transmitted pulse at the emitter transition frequency~\cite{Chen11}. 
This dip is not apparent in the plot of $N(z,t)$ for the case of two initially coincident pulses in Fig.~\ref{sameSide_dynamics_coin}, 
but is present in the two-photon wavepacket in $z$-space as indicated for $t=10.0$. Physically, 
it means that a photon may be detected at a position corresponding to the dip, but \textit{if} the first photon is detected there, 
the probability of detecting the second photon at that 
position is zero, exemplifying that the single-photon scattering features manifest themselves in two-photon scattering, 
although they may not be apparent from the photon density $N(z,t)$. 

To summarise, we have illustrated the full scattering dynamics of two photons on a TLE 
by calculating the total system state at all times. For well-separated uncorrelated single-photon pulses, 
the dynamics may be well approximated by the single-photon results~\cite{Chen11}. As the displacement between the pulses becomes smaller, 
non-trivial dynamics can be induced due to the saturation of the TLE. 
The approach we use here naturally accommodates this regime of two photon scattering. 

\subsection{Transmission and reflection properties}
In order to investigate the transmission properties of the TLE, we consider the relative number of photons propagating to the 
left and right during the scattering process. We can calculate the total number of 
photons propagating to the right as
\beqa
N_\mathrm{R}(t)&=&\int_{0}^\infty \mathrm{d}k\,\bra{\psi(t)}a^\dagger(k)a(k)\ket{\psi(t)} \\
&=&2\int_{0}^\infty \mathrm{d}k\,\int_{-\infty}^\infty \mathrm{d}k'\,|C^\ggg(k,k',t)|^2+\int_{0}^\infty \mathrm{d}k\,|C^\eee(k,t)|^2,
\eeqa
while the total number propagating to the left, 
$N_\mathrm{L}(t)=\int_{-\infty}^0 \mathrm{d}k\,\bra{\psi(t)}a^\dagger(k)a(k)\ket{\psi(t)}$, 
is given by a similar expression with the integration range over $k$ changed to $[-\infty,0]$. The 
excitation probability of the TLE is given by
\beqa
P_\eee(t)=\bra{\psi(t)}c^\dagger c\ket{\psi(t)} = \int_{-\infty}^{\infty}\mathrm{d}k\,\big|C^\eee(k,t)\big|^2,
\eeqa
and normalization of the total state ensures $N_\mathrm{R}(t)+N_\mathrm{L}(t)+P_\eee(t)=2$; 
there is a total of two excitations in the system at all times. We therefore 
define the {\emph{relative}} transmission to the right and left as $T_{\mathrm{R}}(t) = N_\mathrm{R}(t)/2$ and 
$T_{\mathrm{L}}(t) = N_\mathrm{L}(t)/2$.

\begin{figure}
\begin{flushright}
\includegraphics[width=1\textwidth]{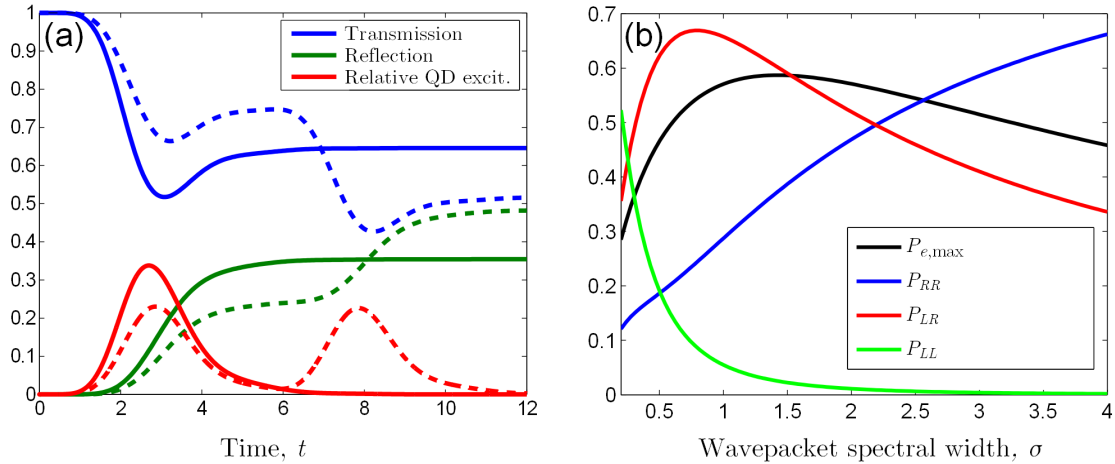}
\caption{\label{streger} (a) Transmission $T_{\mathrm{R}}(t)$ (blue) and reflection $T_{\mathrm{L}}(t)$ (green), together 
with relative TLE excitation, $P_\eee(t)/2$ (red), for parameters corresponding to 
the two cases of perfectly overlapping (solid) and non-overlapping (dashed) pulses shown in 
Figs.~\ref{sameSide_dynamics_coin} and \ref{sameSide_dynamics_sep} respectively. 
(b) Maximum TLE excitation and the directional scattering probabilities 
as a function of the wavepacket $k$-space width, $\sigma$, for two coincident but uncorrelated, single-photon 
pulses with the same width and carrier frequency, resonant with the TLE transition.
}
\end{flushright}
\end{figure}

In Fig.~\ref{streger}(a) we show the left and right transmission coefficients, together with the 
TLE excitation as a function of time, for the two cases of perfectly overlapping (solid) and non-overlapping pulses (dashed) 
introduced in Figs.~\ref{sameSide_dynamics_coin} and \ref{sameSide_dynamics_sep} respectively. 
From these plots a clear reduction in the reflective nature of the TLE when the two pulses are coincident is evident, 
clearly illustrating that the first photon induces partial transparency in the TLE, minimising the interaction between the TLE and the second 
photon.\footnote{Due to the symmetry, the maximum achievable TLE excitation for a single-pulse excitation from a 
single side is 1/2, which is obtained for a pulse with a temporal shape which is exactly the inverse of a pulse emitted by the 
TLE~\cite{Rephaeli10}. Such a pulse would render the TLE completely transparent.} Also evident 
is a temporal delay between excitation of the TLE and the accumulation of the reflected field, 
demonstrating non-instant scattering due to the finite decay rate of the TLE. 

The transmission and reflection coefficients do not contain information regarding scattering-induced 
correlations between the photons, and to that end we define scattering probabilities for the three possible 
directional outcomes of the scattering process. In the long-time limit, the probability that both photons propagate to the right is given by 
\beqa
P_{RR}&=& \frac{1}{2}\lim_{t\rightarrow\infty}\int_{0}^\infty \textrm{d}k\int_{0}^\infty \textrm{d}k'\,\bra{\psi(t)}a^{\dagger}(k)a^{\dagger}(k')a(k')a(k)\ket{\psi(t)} \\
&=& \lim_{t\rightarrow\infty} \int_{0}^\infty \textrm{d}k\int_{0}^\infty \textrm{d}k'\,|C^\ggg(k,k',t)|^2,
\eeqa
while $P_{LL}$ is given by a similar expression with the integration ranges changed to $[-\infty,0]$. 
The probability of having one photon travelling in each of the two directions is
\beqa
P_{LR}(t)&=&2\lim_{t\rightarrow\infty} \int_{-\infty}^0 \textrm{d}k\int_{0}^\infty \textrm{d}k'\,|C^\ggg(k,k',t)|^2.
\eeqa
The scattering probabilities $P_{RR}$, $P_{LR}$, and $P_{LL}$ are thus obtained by integrating the two-photon 
wavepacket over the corresponding quadrant(s) in Fig. \ref{sameSide_dynamics_coin} or Fig. \ref{sameSide_dynamics_sep} in either $z$- or $k$-space.

At long times well past the scattering event, when the TLE has fully decayed to its ground state, the probabilities 
we have defined satisfy $P_{RR}+P_{LL}+P_{LR}=1$. In contrast to the quantities $T_R$ and $T_L$, the 
probabilities $P_{RR}$, $P_{LL}$, and $P_{LR}$ contain 
information regarding the directional correlation between the individual photons. 
The correlations depend crucially on the width of the photon wavepacket, as well as the initial emitter excitation. 
To investigate this, Fig.~\ref{streger}(b) shows the directional scattering 
probabilities as a function of the width of two equal coincident input pulses, together with the maximal emitter excitation, $P_{\eee,\textrm{max}}$. 
The scattering of monochromatic pulses (infinitely small $\sigma$) is well-known~\cite{Auffeves07,Shen09}; all of the pulse 
is reflected when the carrier frequency is resonant with the emitter transition, agreeing with our results here 
in the limit of a small $\sigma$. Here the TLE excitation remains low due to the low optical power in the pulse. 
Spectrally broad pulses have a small overlap with the TLE in $k$-space, resulting in a small degree of interaction and thus 
also a low value of $P_{\eee,\textrm{max}}$ and a high value of  $P_{RR}$. 
The largest $P_{\eee,\textrm{max}}$ is obtained for $\sigma\sim \Gamma/v_{\ggg}$ which is also 
the parameter regime where $P_{LR}$ dominates. This occurs when the spectral overlap between the 
wavepacket of the input state and the TLE emission spectrum is large.

\section{Counter-propagating pulses}
\label{counter}
\begin{figure}
\begin{flushright}
\includegraphics[width=1\textwidth]{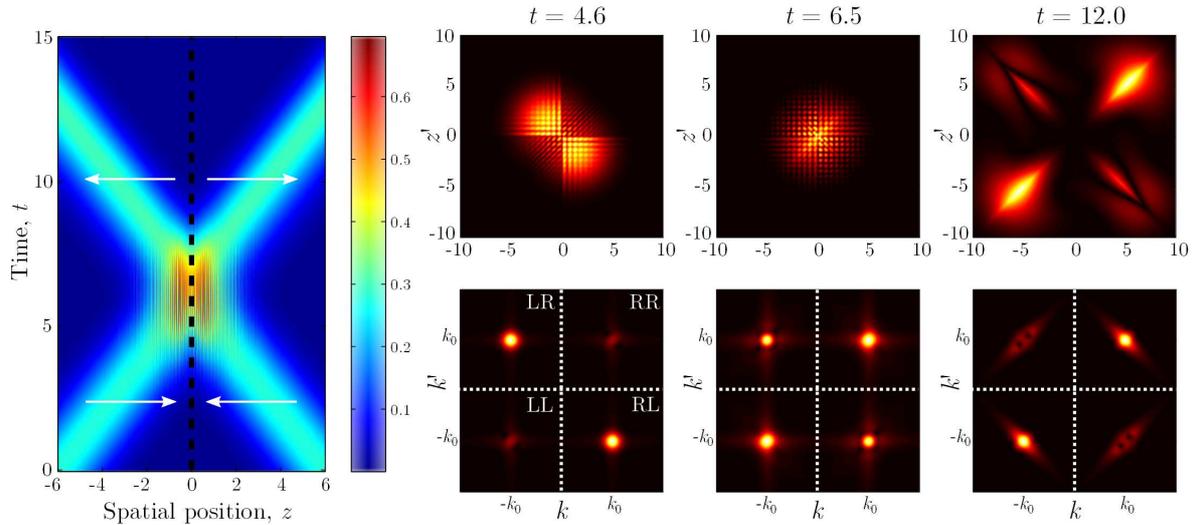}
\caption{\label{diffSide_dynamics} 
Left: Photon density, $N(z,t)$ for an initially uncorrelated ($\sigma_\textrm{p}\rightarrow \infty$) 
two-photon state scattering on the emitter placed at $z=0$, using pulse widths $\sigma_1 =\sigma_2=0.5$ and initial centre positions 
$z_{0,1}=z_{0,2}=-6$. The position of the emitter at $z=0$ is indicated by the dashed black line.
Absolute value of the two-photon wavefunction shown at three different times during the scattering event, 
both in $z$-space (upper row) and $k$-space (lower row). In the $k$-space plots, only the intervals 
centred at $k,k'=\pm k_0$ are shown.
}
\end{flushright}
\end{figure}

We now turn to the case where the TLE is illuminated 
by two counter-propagating single-photon pulses, one photon from each side of the TLE. The corresponding waveguide excitation 
dynamics is shown in Fig.~\ref{diffSide_dynamics}, for excitation pulses with a carrier frequency resonant with the TLE transition energy. 
Due to the symmetry of the scattering problem around $z=0$, the expectation value of the photon density is the same for the left and 
right propagating components of the pulse.  

Closer inspection of the two-photon wavepacket on the right of Fig.~\ref{diffSide_dynamics} reveals 
interesting features regarding the induced correlations. 
We see that $P_{LR}$ is much smaller than $P_{RR}$ and $P_{LL}$. 
This indicates a strong directional correlation between the two scattered photons as the final state suggests both photons 
will be measured propagating in the same direction with high probability.  We note that this property 
cannot be inferred from the photon density plot. 
This phenomenon is analogous to the well-known two-photon interference which gives 
rise to the Hong-Ou-Mandel dip, wherein 
two identical photons impinging from opposite sides of an optical beam-splitter coalesce and are measured in the same 
output arm~\cite{Hong87}. In the present case, 
however, the effect is only partial due to the non-zero spectral width of the input pulses and the TLE saturation, 
and as a consequence $P_{LR}$ is not zero. 
This beam splitter-like effect has been observed in Ref.~\cite{Longo12} for coupled optical waveguides 
described by a tight-binding model between the individual sites.

\subsection{Induced Correlations}
\label{entanglement}
We now turn our attention to the correlations induced in the two-photon-state 
as a result of the scattering process. First, it is important to establish in which degrees of freedom the photons can be correlated. 
We distinguish between two correlation types, which we refer to as `directional' 
and `modal'. Directional correlations are those present in measurement statistics acquired from 
detecting the \emph{direction of propagation} of each of the two photons, and are captured by the quantities $P_{ij}$ 
for $\{i,j\}\in\{\mathrm{R},\mathrm{L}\}$. If the propagation direction of one photon depends on the measured 
propagation direction of the other, the two are said to have directional correlations. 
Modal correlations, on the other hand, are concerned with measurement statistics obtained when detecting 
the \emph{position} of each photon, assuming a given configuration of propagation directions. 
These modal correlations are contained in the correlation parameter $\sigma_\textrm{p}$, 
defined for the input state in Eq.~(\ref{eq:betazz}). Modal correlations are more 
traditionally described in terms of the well-known second order $g^{(2)}$ correlation function~\cite{Zheng12}, which 
is typically employed when describing intensity correlations. 
A generic two-photon state may be correlated according to one of these measures, but fully uncorrelated in the other. 
Fig.~\ref{rho_plots}(c) shows an example of such a state. The elliptical shape of the wavepacket 
in real-space is a signature of modal correlations,
but the state can have no directional correlations, since both photons are propagating to the right.

The scattering of co-propagating photons shown in Fig.~\ref{diffSide_dynamics} induces strong directional correlations.  
Modal correlations are also induced, as can be seen from the elliptical shape of the 
wavefunction in $z$-space and $k$-space, meaning that the emitted photons are anti-correlated in $k$-space and correlated in $z$-space. 
This can be further appreciated by comparison with the state shown in Fig.~\ref{rho_plots}(c), which was defined to have modal correlations. 
The induced anti-correlation in $k$-space can be understood as a four-wave mixing process, 
where elastic scattering of two photons of almost identical energy results in one photon with higher energy and one with lower energy. 
This gives rise to the elliptical shape of the wavefunction in $k$-space, cf. the spectrum in Fig.~\ref{diffSide_dynamics} at $t=12.0$. 
The correlation in $z$-space implies a larger probability of detecting the second photon spatially close to the first, i.e. photon bunching. 
Modal correlations such as these are not present in the scattered state from a conventional linear optical beam splitter; the 
modal entanglement observed here is caused by a non-linear scattering process between the incoming and emitted photons, 
which is mediated by the excitation of the TLE. 

\subsection{Induced entanglement}
In order to relate the induced quantum correlations in the photonic state to the 
TLE excitation dynamics, we require a measure of the induced correlations, which can be facilitated by entanglement theory. 
There are several proposals in the literature of how to quantify the degree of entanglement (quantum correlations) 
between individual subsystems~\cite{Vidal02,Plenio07,Vedral98}, particularly for distinguishable systems each 
of which may be in one of only two states, 
e.g. two spatially separated spin-half particles. 
These measures include the fidelity, the concurrence, the negativity, and the entropy of entanglement~\cite{Kok_book}, 
each of which has a different operational meaning, and may be more or less appropriate given the problem at hand. 
For two indistinguishable particles e.g. two bosons  
in the same two-photon Hilbert space, extensions to the distinguishable case have to be made~\cite{Wiseman03,Ghirardi04,Eckert02}. 
If the indistinguishable bosons can each occupy more than two states, as is the case for the state expressed by Eq.~(\ref{twofot_1}) 
(where the number of states is equal to the dimension of each particle sub-Hilbert space), there are fewer 
ways to quantify the entanglement. Among these is the von Neumann entropy of the reduced 
single-particle density matrix~\cite{Ghirardi04,Paskauskas01}, 
which quantifies the modal entanglement by 
the degree to which the state of the second photon is affected by a $k$-space measurement on the first. 

In order to explore the extent to which our system behaves as a beam-splitter, we quantify the amount of directional entanglement present in the scattered state. To do this, the two-photon 
state may be projected onto a two-dimensional Hilbert space with each photon being in either a left or a right propagating state, 
giving three basis states, $\ket{LL}$, $\ket{LR}$, and $\ket{RR}$. 
This projected system is identical to the case of two indistinguishable two-state particles discussed above, for which  
the entanglement may be quantified by the fidelity, i.e. by comparison to a maximally entangled state. 
We focus here only on entangled states with a different number of particles in each direction and thus compare to two of the four Bell-states only,
\beqa
\ket{\Phi^\pm}=\frac{1}{\sqrt{2}}\left[\ket{LL}\pm\ket{RR}\right].
\eeqa
For pure states as in Eq.~(\ref{twofot_1}), the fidelities with respect to the maximally entangled states are 
defined as the overlap between the scattered state and the maximally entangled state, $F^\pm=|\braket{\Phi^\pm}{\psi}|^2$ \cite{Kok_book}. 
The fidelities exceed 1/2 only if $\ket{\psi}$ is a non-classical state, and can therefore be 
interpreted as a measure of the directional entanglement.

\begin{figure}[htb]
\begin{center}
\includegraphics[width=18pc]{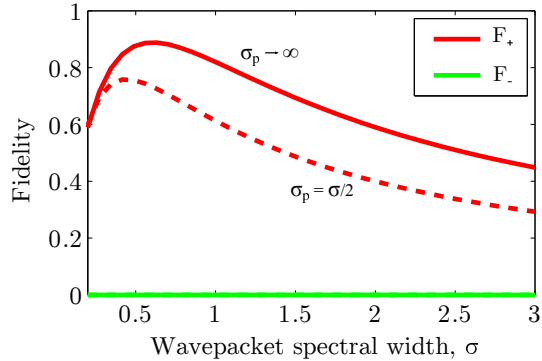}
\caption{\label{entanglement}
Directional entanglement, quantified by the fidelity, plotted versus the spectral width of the photon wavepacket. 
Here we consider two identical, single-photon wavepackets impinging on the TLE from each side, initially equidistant from the 
TLE, and both being resonant with the TLE transition 
(i.e. the conditions are the same as those of Fig.~\ref{diffSide_dynamics} for which $\sigma_1=\sigma_2=0.5$ and $\sigma_p\to\infty$).
Cases of initially uncorrelated states, $\sigma_\textrm{p}\rightarrow \infty$, and spatially correlated states, $\sigma_\textrm{p}=\sigma_1/2$ are shown, 
and the wavepacket widths are always equal $\sigma_1=\sigma_2=\sigma$.}
\end{center}
\end{figure}

For two identical photons scattering on the TLE from each side, as in Fig.~\ref{diffSide_dynamics}, the input state has $F^\pm=0$, 
as the overlap with the initial state $\ket{LR}$ is zero. The fidelity for the scattered state is shown in Fig.~\ref{entanglement} for varying widths of the input pulses. 
The correlated input state, where both photons have a larger probability of scattering on the TLE at the same time, 
leads to a smaller fidelity at the output than for two uncorrelated photons at the input. 
For the initially correlated states, such as that shown in in Fig.~\ref{rho_plots}(c), 
the spectrum of the photons is tighter than the uncorrelated case in Fig.~\ref{rho_plots}(a), 
but the spatial distribution is broadened, resulting in a lower probability of having both of the photons at the 
TLE at the same time; this decreases the induced correlations and correspondingly leads to a smaller fidelity..

In the limit of large $\sigma$, only a small fraction of the pulse interacts with the TLE, 
giving a fidelity which approaches zero. In the small $\sigma$ limit, 
the incoming pulse is temporally broad, resulting in a low light intensity at the TLE position at all times, 
and hence, to a good approximation, the TLE remains in its ground state. As the TLE only induces non-linearities when it is excited, a two-photon packet with 
small $\sigma$ scatters as if the two photons were scattering individually on the TLE, giving the scattered state $\ket{LR}$. 
We note that the maximum fidelity is obtained when the linewidth of the Gaussian input pulses is comparable with the decay rate 
of the TLE. In this case excitation of the TLE is high, and a highly directionally entangled state is produced.

\section{Conclusion}
\label{conclusion}
In conclusion, we have developed a wavefunction approach to study the scattering of two photons on a 
two-level emitter in a one-dimensional waveguide. 
Our method benefits from the simple mathematical form, and provides the full temporal dynamics 
of the scattering event, as well as a detailed description of the scattering-induced correlations. 
For co-propagating pulses, we saw that the excitation 
of the emitter strongly influences its transparency. This results in transmission and reflection coefficients 
which depend sensitively on the separation between the two input pulses. 
For counter-propagating pulses, the emitter--waveguide system shows beam-splitter like features, generating directional correlations 
and entanglement in the 
scattered two-photon state, occurring most strongly when the emitter excitation is largest. Unlike a conventional linear optical beam-splitter, however, the finite 
decay rate of the emitter introduces non-linearities which manifest as additional bunching effects. 
Finally, we note that our model could be extended to more complicated scattering scenarios, 
such as several quantum dots with possibly additional levels~\cite{Witthaut10,Martens13}. The numerical approach we use also allows for the investigation of the role of waveguide dispersion, as well as non-Markovian 
coupling to the scattering object by including frequency-dependent coupling coefficients in the system, which we plan to investigate in future work.


\section{Acknowledgments}
This work was supported by Villum Fonden via the Centre
of Excellence “NATEC”, the Danish Council for Independent Research (FTP 10-093651) and by the European Metrology Research
Programme (EMRP) via the project SIQUTE (Contract
No. EXL02). The EMRP is jointly funded by the EMRP participating countries within EURAMET and the European Union.

\end{document}